\documentclass[sigconf]{acmart}

\newcommand{\etal}{\emph{et al}.}

\AtBeginDocument{
  \providecommand\BibTeX{{
    \normalfont B\kern-0.5em{\scshape i\kern-0.25em b}\kern-0.8em\TeX}}}

\copyrightyear{2024} 
\acmYear{2024} 
\setcopyright{rightsretained} 
\acmConference[UIST '24]{The 37th Annual ACM Symposium on User Interface Software and Technology}{October 13--16, 2024}{Pittsburgh, PA, USA}
\acmBooktitle{The 37th Annual ACM Symposium on User Interface Software and Technology (UIST '24), October 13--16, 2024, Pittsburgh, PA, USA}
\acmDOI{10.1145/3654777.3676412}
\acmISBN{979-8-4007-0628-8/24/10}

\usepackage{amsmath} 
\usepackage{multirow}
\usepackage{enumerate}
\usepackage[linesnumbered, ruled]{algorithm2e}
\SetKwInput{KwHyper}{Hyperparameters}

\usepackage{comment}

\begin{document}

\title[V-Hands: Touchscreen-based Hand Tracking for Remote Whiteboard Interaction]{V-Hands: Touchscreen-based Hand Tracking for Remote Whiteboard Interaction}

\author{Xinshuang Liu}
\email{xinshuangliu@outlook.com}
\additionalaffiliation{
    \institution{Microsoft Research Asia}
    \city{Beijing}
    \country{China}
}
\affiliation{
  \institution{UC San Diego}
  \country{USA}
}
\authornote{Work done during the internship at Microsoft Research Asia.}

\author{Yizhong Zhang}
\authornote{Corresponding author.}
\email{yizzhan@microsoft.com}
\affiliation{
  \institution{Microsoft Research Asia}
  \country{China}
}

\author{Xin Tong}
\email{xtong.gfx@gmail.com}
\affiliation{
  \institution{Microsoft Research Asia}
  \country{China}
}

\begin{abstract}
  In whiteboard-based remote communication, the seamless integration of drawn content and hand-screen interactions is essential for an immersive user experience. Previous methods either require bulky device setups for capturing hand gestures or fail to accurately track the hand poses from capacitive images. In this paper, we present a real-time method for precise tracking 3D poses of both hands from capacitive video frames. To this end, we develop a deep neural network to identify hands and infer hand joint positions from capacitive frames, and then recover 3D hand poses from the hand-joint positions via a constrained inverse kinematic solver. Additionally, we design a device setup for capturing high-quality hand-screen interaction data and obtained a more accurate synchronized capacitive video and hand pose dataset. Our method improves the accuracy and stability of 3D hand tracking for capacitive frames while maintaining a compact device setup for remote communication. We validate our scheme design and its superior performance on 3D hand pose tracking and demonstrate the effectiveness of our method in whiteboard-based remote communication. 
Our code, model, and dataset are available at \textcolor{blue}{\texttt{\href{https://V-Hands.github.io}{https://V-Hands.github.io}}}.

\end{abstract}



\begin{teaserfigure}
\begin{center}
    \centering
    \includegraphics[width=0.9\textwidth]{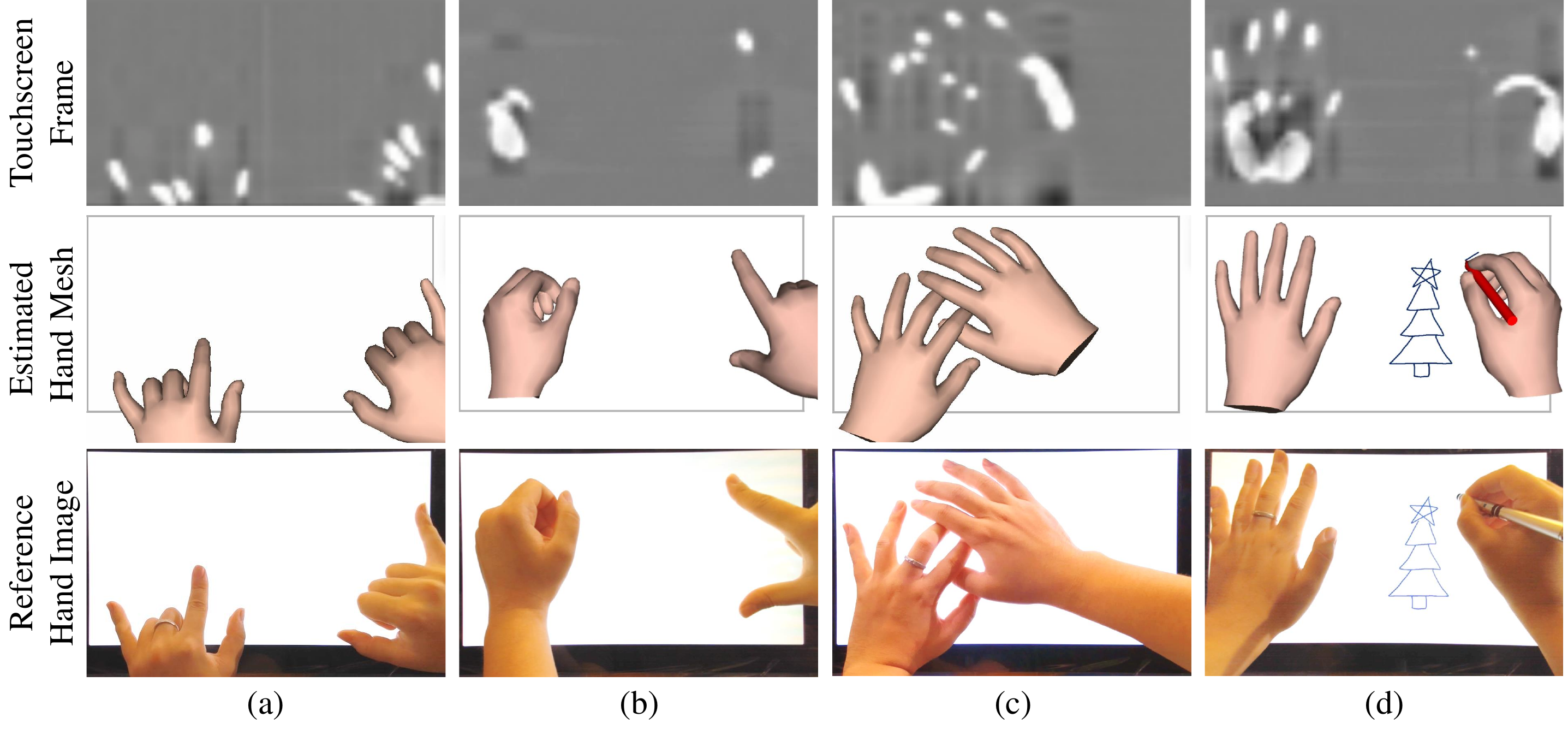}
    \caption{Given the capacitive frames of two hands of a subject interacting with the touchscreen (first row), our method reconstructs the corresponding 3D hand poses (second row) in real time. The ground truth hand poses of the captured capacitive frames are displayed in the third row for comparison. Note that our method can accurately reconstruct various poses of two hands, even for overlapped hands as demonstrated in (c).
    }
    \label{fig:teaser}
\end{center}
\end{teaserfigure}

\maketitle

\section{Introduction} \label{Sec:Introduction}

Capacitive touchscreens enable users to naturally interact with displayed content via hand fingers, making them widely used as virtual whiteboards in remote communications for efficient user interaction. However, unlike on-site collaborations where participants can simultaneously perceive both the user's gestures and the content displayed on the screen, a local user cannot observe the hand poses of other remote users, which significantly degrades the interaction efficiency and immersive experience of remote communications ~\cite{bai2020user}.

A set of methods have been proposed to address this issue. Camera-based solutions ~\cite{bai2020user, thily2010using, Le2017, Genest2013, Sasa2012, Tang2007, Iwai2018, Coldefy2007, Izadi2007} employ additional cameras installed on the top of a touchscreen to capture and reconstruct hand poses. While these methods can accurately track hand poses in real time, the cost and size of the extra device setups limit their accessibility and portability, restricting their use in many remote communication scenarios.

Touchscreen-based solutions ~\cite{DBLP:conf/mhci/ChoiM021,DBLP:conf/uist/AhujaS021} directly infer hand poses from capacitive images of the touchscreen. On one hand, this approach avoids extra device setup and preserves the accessibility and portability of the device for efficient user interaction. On the other hand, due to the limited range and resolution of the capacitive sensors and the occlusion between different hand parts, the capacitive image only covers a small part of the hands (as shown in Fig.~\ref{fig:teaser}), making reconstructing the hand poses from these incomplete capacitive images a challenging problem. Existing methods can only recover static hand poses of single-hand gestures with limited accuracy, hindering their use in real applications where the poses of two hands continuously change during the interaction between hand and screen.  

In this paper, we present a method for accurately tracking 3D hand poses from capacitive video frames in real time. Our method estimates continuous 3D poses of two hands interacting with the screen and addresses complex hand gestures as shown in Fig.~\ref{fig:teaser}. To achieve this, our method first estimates 3D hand joint positions from input capacitive frames using a deep neural network, then reconstructs hand poses with a constrained inverse kinematic (IK) solver. For the joint position estimator, we adapt an RGB-video-based 3D hand pose estimation network to capacitive image input. The network output includes hand classification results (i.e., left or right) and 3D positions of hand joints. We further incorporate a gated recurrent unit (GRU) architecture into the network, which exploits the temporal information to resolve the ambiguity in hand classification and pose estimation. Finally, the constrained IK solver takes the 3D hand joint positions as input and infers 3D transformations of hand joints for animating a template hand mesh to the resulting 3D hand poses. 

For network training, we develop a data acquisition system for capturing capacitive video frames and corresponding 3D hand skeletons from multi-view RGB images and obtain a new dataset. Compared to existing datasets that only consist of static capacitive images and inaccurate hand poses ~\cite{DBLP:conf/uist/AhujaS021}, our new dataset is composed of sequences of two-hand gestures with higher pose accuracy and more diverse hand poses.  

We evaluate the accuracy of our method using capacitive images of different hand poses and demonstrate its advantages over existing solutions. We also assess the effectiveness of our method in two remote communication applications. Experimental results indicate that our system enables users to focus more on interaction content, thus improving interaction efficiency. We will release our dataset, code, and model on our project web page.

In summary, our work makes the following contributions:
\begin{itemize}

\item We introduce a real-time 3D hand tracking method from capacitive frames that provides temporally consistent and accurate 3D poses of two hands when interacting with a touchscreen.

\item We introduce the first capacitive video dataset for 3D hand tracking, which consists of capacitive video clips and corresponding accurate 3D hand poses. The dataset enables our method and facilitates future research on capacitive-based hand pose estimation.

\item We apply our method in two whiteboard-based remote communication applications on the touchscreen, which efficiently enhances the user experience while preserving the portability of the device.

\end{itemize}

\section{Related Work} \label{Sec:Related Work}

Our work is related to touchscreen technology, hand pose estimation and remote whiteboard interactions. Below we review related work in these areas.

\subsection{Touchscreen and Touch Sensing Technology}

Touchscreen technology provides an integrated and intuitive solution that seamlessly combines display and user input. 
The history of touchscreens can be traced back to the pioneering work of E.A. Johnson ~\cite{johnson1965touch} in 1965. Johnson devised an innovative way to place an array of capacitive sensitive electrodes on the surface of the CRT screen, effectively creating the basis for finger-actuated touchscreens. 
Over the subsequent decades, a variety of touch sensing technologies were invented for different application scenarios, including capacitive ~\cite{dietz2001diamondtouch, rekimoto2002smartskin}, resistive ~\cite{hurst1970touch}, acoustic wave ~\cite{nara2001surface}, optical ~\cite{han2005low} and dense optical sensor array ~\cite{izadi2009thinsight}.
R.A. Boie ~\cite{boie1984capacitive} introduced the first multi-touch screen utilizing capacitive sensing in 1984. 
Further enhancing the function of capacitive touchscreens, Wayne Westerman ~\cite{westerman1999hand} developed algorithms to accurately detect multi-touch points on the screen. This technology was subsequently acquired by Apple Inc. and gained widespread adoption with the release of the iPhone in 2007. 
Capacitive touchscreens offer many advantages over other methods, such as multi-touch support, superior optical quality, and enhanced durability ~\cite{DBLP:journals/sensors/NamSLCJ21}. As a result, they have become the industry standard for mobile phones, tablets, and laptops, fundamentally changing the way we interact with digital devices.

In contemporary consumer devices, capacitive touchscreens use integrated circuits with embedded touch detection algorithms to achieve low-latency multi-touch prediction ~\cite{Atmel2015maXTouch}. To ensure accuracy and efficiency, these algorithms are specifically designed to detect isolated fingertips. Therefore, if two fingers are too close together or the touch area is too large, the touch point may not be detected accurately, thereby limiting the range of freestyle interaction with the touchscreen. 
With the rapid development of machine learning techniques, data-driven methods have become an important tool to enhance raw touchscreen data processing. These methods help improve the resolution of touch image sensing ~\cite{mayer2021super, streli2021capcontact} and further recover detailed hand poses, including finger recognition ~\cite{le2019investigating, tung2018raincheck, huang2024specifingers}, touch direction estimation ~\cite{DBLP:conf/chi/RogersWSM11, DBLP:conf/tabletop/XiaoSH15, DBLP:conf/tabletop/MayerLH17}, gesture recognition ~\cite{le2018palmtouch, DBLP:conf/mc/SchweigertLHWLM19}, pressure estimation ~\cite{DBLP:conf/mhci/BoceckSLM19}, or even some other objects placed on the touchscreen ~\cite{schmitz2021itsy}, thus offering the potential to extend the capabilities of touchscreen interactions. 
Compared to these methods that focus on improving touch sensing or enhancing captured signals, our method aims to infer the invisible 3D hand poses from the captured capacitive images.

\subsection{Hand Pose Estimation}

3D hand pose estimation and tracking from RGB \cite{DBLP:conf/cvpr/BoukhaymaBT19,DBLP:conf/cvpr/BaekKK19,DBLP:conf/cvpr/GeRLXWCY19,DBLP:conf/wacv/LinWM21,DBLP:conf/eccv/MengJLQLOL22} or depth images \cite{DBLP:conf/iccv/XiongZ0CYZY19,DBLP:journals/tog/MuellerDBSVOCT19,DBLP:conf/eccv/FangLLXK20,DBLP:conf/cvpr/RenSHW0L22,DBLP:conf/aaai/ChengWZMGTW0022} have been extensively studied in the past decades, with remarkable progress achieved through the introduction of deep neural networks.
Recent advances in computer vision have inspired novel methods in this domain, such as adopting the Transformer architecture \cite{DBLP:conf/nips/VaswaniSPUJGKP17} in model design \cite{jiang2023a2j,wen2023hierarchical,DBLP:journals/sensors/KanisGKBSH23}, enhancing models with vision-language pre-trained models \cite{DBLP:conf/wacv/LeePKKBB23}, and training the models using contrastive learning \cite{spurr2021self,lin2023cross}. In addition to methodological improvements, the emergence of new datasets \cite{DBLP:conf/accv/MyanganbayarMDK18,DBLP:conf/iccv/ZimmermannCYRAB19,DBLP:journals/ivc/Gomez-DonosoOC19,DBLP:conf/eccv/MoonYWSL20,DBLP:conf/eccv/BrahmbhattTTKH20} has further enhanced model performance in hand pose estimation. 
However, challenges still exist in estimating hand poses with limited information ~\cite{DBLP:journals/sivp/LinLZSM22,DBLP:conf/wacv/WangML23,xu2023h2onet}. Recent research has started exploring the use of parametric hand models ~\cite{DBLP:journals/tog/0002TB17} to tackle issues with incomplete data, such as low resolution or occlusions ~\cite{DBLP:conf/cvpr/BoukhaymaBT19}. While these methods can predict reasonable hand poses for occluded hand parts, the accuracy of visible parts may be compromised due to the reduced motion space. 

Different from the RGB or depth images that capture the shape and appearance of full hands, capacitive images only record a low-resolution grayscale image of the partial hand regions that touch the screen, which is similar to the pressure-sensing floor in ~\cite{branzel2013gravityspace}. As a result, estimating the invisible hand poses from capacitive images is a much more challenging task.
Chung \etal ~\cite{DBLP:conf/eurographics/ChungKHP15} proposed a method for estimating a hand model from touch points, which depends on pre-detected touch points and is therefore unsuitable for interaction with free hand gestures. Le ~\etal ~\cite{le2018infinitouch} utilized a convolutional neural network (CNN) to identify individual fingertips directly from raw capacitive touch images. Recent research has sought to estimate gestures directly from incomplete touchscreen data. 
Choi \etal ~\cite{DBLP:conf/mhci/ChoiM021} constructed a hand pose database and subsequently matched the touchscreen image to one of the reference hand poses, employing nonlinear deformation to align the matched reference hand pose with the touchscreen image. However, due to the complexity of hand gestures and the continuous nature of hand movement, predefined hand categories may not cover all possible hand gestures.
Ahuja \etal ~\cite{DBLP:conf/uist/AhujaS021} improved this result by developing a CNN encoder and a multi-layer perceptron (MLP) to estimate joint poses from each touch frame, but they did not take advantage of the temporal consistency of frames, and the direct prediction of Cartesian coordinates of joints difficult to precisely align the touch points on the touch image.
Moreover, these approaches do not accommodate scenarios in which both hands interact with the touchscreen simultaneously. Our work introduces a novel approach for accurate hand tracking on capacitive touchscreens, addressing the limitations by enabling interaction with free hand gestures, using heatmap representation for precise touch point alignment (as evidenced by improved accuracy in recent studies ~\cite{qu2022heatmap, luo2021rethinking, yang2020seqhand}), and support for dual-hand scenarios. 

Inverse Kinematics (IK) also plays a crucial role in hand pose estimation by recovering the hand skeleton and hand mesh from joint positions. It involves fitting joints to target positions while adhering to the connectivity and geometric constraints of the template hand skeleton ~\cite{DBLP:journals/tsmc/Wampler86, DBLP:journals/jgtools/BussK05, DBLP:conf/icra/Pechev08, DBLP:journals/cvgip/AristidouL11}. Once the hand skeleton is obtained, the hand mesh can be bound to the skeleton using automatic rigging techniques ~\cite{baran2007automatic} and deformed with linear blend skinning (LBS) ~\cite{lewis2000pose}. 
In our approach, we employ an IK solver with constraints tailored for touchscreen interactions. For hand mesh generation, we utilize the MANO hand model \cite{DBLP:journals/tog/0002TB17}, which provides predefined rigging weights learned from data for a realistic representation of the user's hand. By combining IK, LBS, and MANO hand models, our approach effectively recovers hand bones and hand meshes from the predicted joint positions.

\subsection{Remote Whiteboard Interaction}

Our work is also related to remote whiteboard interaction, specifically enhancing user experience by integrating visual cues of hand gestures.
Previous work has validated and demonstrated that displaying hand movements can efficiently enhance remote whiteboard interactions and facilitate remote collaboration ~\cite{bai2020user}.
Remote whiteboard has two different settings: typical whiteboard settings  ~\cite{thily2010using} and lightboard settings ~\cite{mccorkle2020lightboard}. In a typical whiteboard setting, each user is given a horizontal desktop display showing shared working items. To convey visual cues of the hand gesture, a vertically downward-facing camera is installed above the desktop display to capture the local user's hand movements and transmit the segmented hand image to the remote user in real time ~\cite{Le2017, Genest2013, Sasa2012, Tang2007}. Some systems also include a separate vertical screen to display the remote user's portrait video ~\cite{Iwai2018, unver2016sharetable, Coldefy2007, Izadi2007}.

Lightboard settings represent another category of remote collaboration setups. In this configuration, the vertical screen displays working items while also displaying a life-size portrait of the remote user, creating the illusion of collaborative work with users on either side on a clear glass surface. Some works use cameras placed behind transparent screens to capture hand movements ~\cite{Tang91, Ishii1992, Wilson04touchlight, Tan2009}; however, this approach requires specialized hardware and takes up a lot of space to install. An alternative approach is to synthesize a 3D hand model using hand tracking sensors mounted around or in front of the screen ~\cite{Zillner2014, Wood2016, zhang2023remotetouch}.

Compared to existing solutions, our approach eliminates the need for assistive devices to provide visual cues for gestures during remote collaboration and supports both whiteboard and lightboard configurations by leveraging touch screens for hand tracking. Our approach provides a more versatile and simplified solution with the potential to be implemented on a variety of business mobile devices, thereby enhancing remote collaboration environments.

\begin{figure*}
    \centering
    \includegraphics[width=\textwidth]{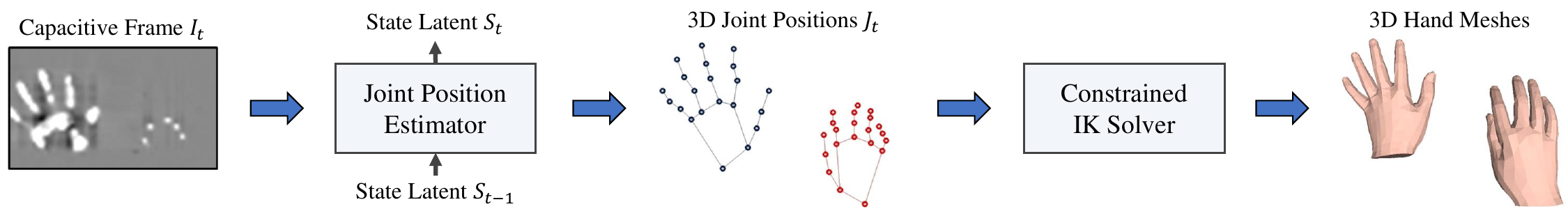}
    \caption{An overview of our method. Given the current capacitive frame $I_{t}$ and a state latent $S_{t-1}$ aggregated from previous frames, our joint position estimator infers the 3D joint positions $J_{t}$ of the two hands in the current frame and subsequently updates the state latent to $S_{t}$. Following this, a constrained inverse kinematic (IK) solver is employed to reconstruct the 3D hand pose from the 3D hand joint positions and subsequently transform the 3D hand meshes to the current pose.}
    \label{fig:overview}
\end{figure*}

\section{Method} \label{Sec:Implementation}

Fig.~\ref{fig:overview} provides an overview of our proposed method, which sequentially processes each touch frame as input and generates estimated hand meshes as output. Our approach consists of two key components. The first is a joint position estimator that estimates each joint position for both hands in 3D space. The hand pose history is embedded as a hidden state within the GRU unit of the estimator, enabling temporal consistent prediction of joints not in contact with the touchscreen. The second component is a constrained IK solver that computes parameters for generating hand mesh using the estimated joints. The constrained IK solver also takes into account previous results to minimize noise and ensure smooth hand motion.

\subsection{Joint Position Estimator}

We use a neural network to estimate the joint position of each hand in the touch image, as illustrated in Fig.~\ref{fig:model}. For a given touch image of the current frame, the algorithm first determines the existence of each hand within the image and subsequently predicts the 3D coordinates of each joint. To enhance accuracy, we employ a heatmap and depth distribution for each joint, which has been demonstrated to be more accurate than directly predicting the coordinate values \cite{DBLP:conf/eccv/SunXWLW18}.

\subsubsection{Preprocessing}

The touchscreen has a rectangular area of $345mm \times 195mm$ and captures skin contact using a capacitance map. This map is represented as a $71 \times 41$ 8-bit grayscale image. To process the image, we first normalize it to a range of $[0-1]$. When there is no contact, the pixel value is theoretically 0.5, while skin contact increases the value to 1.0. However, factors like fingerprints and smudges can cause variations in these values. Based on our observations, we filter out noise and separate contacting and non-contacting regions by clamping pixel values in the range of $[0.0, 0.6]$ to 0.0. This allows us to identify areas where the skin touches the screen accurately.

As users may touch the screen freely, the projection of their joints might fall outside the image range. To account for this, we expand the image to  $128 \times 96$ pixels (with a physical range of $620mm \times 456mm$) by adding padding around the original image. We also include an additional channel to distinguish between original and extended pixels. This expansion allows for accurate representation of joint positions using heatmaps regardless of how the hand touches the screen and simplifies the upscaling and downscaling of the image.

\subsubsection{Network Structure}

\begin{figure*}
    \centering
    \includegraphics[width=\textwidth]{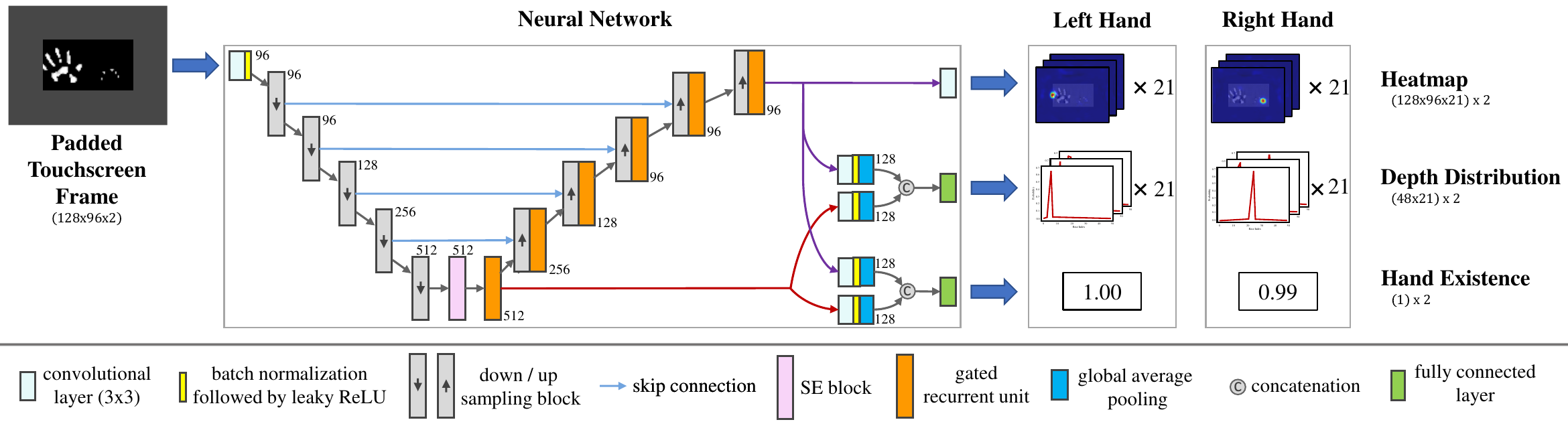}
    \caption{The network architecture of the joint position estimator.}
    \label{fig:model}
\end{figure*}

Our model is based on U-Net architecture ~\cite{ronneberger2015u, steuerlein2022conductive} with additional recurrent modules, similar to the one described in ~\cite{DBLP:journals/spic/ChenGPB22}  (see Fig.~\ref{fig:model}).
The model encodes a two-channel, $128\times96$ touch image into a $4\times3\times512$ latent code using five downsampling blocks. The latent code is then refined by a Squeeze-and-Excitation (SE) block \cite{DBLP:conf/cvpr/HuSS18}, which dynamically recalibrates channel-wise features.

The heatmaps, with dimensions of $128\times96\times42$, are retrieved through five upsampling blocks and a convolution layer. Each upsampling block is followed by a gated recurrent unit (GRU) to record the latent state of previous frames. The heatmaps have 21 channels for each hand, indicating the position of each joint. Further details about the downsampling and upsampling blocks can be found in the Appendix.

The joint projections on the screen are predicted using heatmaps, while the distance of each joint to the screen is predicted separately. The distance is represented by a depth distribution within the range of -1cm to 11cm relative to the screen, which covers the space of human hand interactions with the screen. The range is evenly divided into 48 sections, giving a resolution of $2.5mm$. In our implementation, the final GRU output and latent code are transformed into a 128-element vector using a convolution layer, followed by batch normalization, leaky ReLU activation, and a global average pooling layer. These vectors are concatenated, and the depth distribution is predicted using a fully connected layer with subsequent softmax activation for normalization.

Using the same network architecture as that for depth distribution estimation (only modifying the final activation layer from softmax to sigmoid), we detect the existence of each hand and quantify it as two real numbers within $[0, 1]$.

\subsubsection{Loss function}

Our loss function

\begin{equation}
    \label{eq:loss_final}
    L = \lambda_H L_H + \lambda_D L_D + \lambda_B L_B + \lambda_E L_E,
\end{equation}

consists of four components $L_H$, $L_D$, $L_B$, and $L_E$.

\begin{equation}
    \label{eq:loss_heat}
    L_H = ||H - H_{gt}||_2,
\end{equation}

supervises the heatmap prediction of hand joints. Here, $H$ represents the predicted heatmaps, and $H_{gt}$ represents the ground truth heatmaps.

\begin{equation}
    \label{eq:loss_depth}
    L_D = ||d - d_{gt}||_2,
\end{equation}

supervises the distance of predicted hand joints to the screen. In this case, $d$ refers to the predicted depth distribution of hand joints, and $d_{gt}$ refers to the ground truth.

\begin{equation}
    \label{eq:loss_bone}
    L_B = ||l - l_{gt}||_2,
\end{equation}

supervises the bone length of each hand skeleton to achieve a reasonable hand pose. Here, $l$ is the predicted bone length, and $l_{gt}$ is the ground truth value. 

\begin{equation}
    \label{eq:loss_existence}
    L_E = BCE(E, E_{gt}),
\end{equation}

supervises the existence of each hand in the image. $BCE$ is the binary cross-entropy loss function, $E$ is the value indicating the existence of a hand, and $E_{gt}$ is the ground truth value. 

In our experiments, we set the weight parameters $\lambda_H=10$, $\lambda_D=2$, $\lambda_B=1$, and $\lambda_E=0.2$. These weights balance the contribution of each component in the overall loss function.

\subsubsection{Training}

In our training process, we take advantage of the symmetry between the left and right hands by performing data augmentation. This is done by flipping the image horizontally and swapping the positions of the left and right hands. We use the Adam optimizer \cite{DBLP:journals/corr/KingmaB14} to train our model for 1000 epochs, confirming convergence without overfitting by observing the loss curve.

During each epoch, we randomly select N consecutive frames from each touchscreen video as input data. The value of N starts at 2 and increases by 1 after every 20 epochs until it reaches 30. We initialize the learning rate at $1x10^{-3}$ and multiply it by 0.999 every epoch throughout the training process.

\subsubsection{Joint Estimation}

In this joint estimation algorithm, we perform test-time augmentation to enhance the accuracy of our predictions. This is achieved by inferring both the original touch image and its flipped version. The predicted results are then combined using an averaging technique. 

The next step is to obtain the 3D coordinates of each joint from the combined heatmaps and depth distributions. We adopt a classification-regression scheme for coordinates estimation to achieve better accuracy and robustness ~\cite{pintea2023step}. For each joint, its 3D coordinate $(\hat{x}, \hat{y}, \hat{z})$ is calculated using the following equation:

\begin{equation}  
    \left\{  
    \begin{array}{@{}l}  
        \hat{x}, \hat{y} = \mathop{\arg\max}\limits_{x, y} \mathbf{H}_{x,y} \\  
        \hat{z} = \mathbf{w}^T \mathbf{d},  
    \end{array}  
    \right.  
\end{equation}  
where $\bf{H}$ represents the heatmap of the corresponding joint, $\mathbf{d}$ denotes the 48 basis of depth uniformly sampled in the range of $[-1cm, 11cm]$, and $\bf{w}$ is the predicted depth distribution. The existence of each hand is determined by the hand existence coefficient. If the coefficient is greater than 0.5, it indicates that the corresponding hand is detected in the touch image.

\subsection{Constrained Inverse Kinematics Solver}

In this section, we aim to predict the hand skeleton and mesh by solving an inverse kinematics problem. The input to this process is the 3D coordinates of each individual joint, which are the output of the hand joint estimator. To maintain a natural hand pose and obtain the hand shape, we apply constrained inverse kinematics to the MANO hand model \cite{DBLP:journals/tog/0002TB17}. This approach differs from parametric hand models \cite{DBLP:journals/tog/0002TB17, DBLP:conf/cvpr/BoukhaymaBT19}, as it operates directly on the joints, resulting in better touch accuracy.

The MANO model of each hand consists of two sets of parameters: pose parameters $\mathbf{\Theta}\in\mathbb{R}^{48}$, which represent the rotation angles of each joint in root centered coordinate to control the hand skeleton and deform the hand mesh using linear blend skinning, and shape parameters $\mathbf{\beta}\in\mathbb{R}^{10}$, which reconstruct the hand shape in the rest state. 
Based on the anatomical structure of the real hand, we confine the pose parameters of each hand as $\mathbf{\theta}\in\mathbb{R}^{26}$, including 6 DoFs for root, 2 DoFs for CMC joint of the thumb, 2 DoFs for MCP joints of other fingers and 1 DoF for all other 10 joints.
We denote the joints of the MANO hand model as $\mathbf{f}(\mathbf{\theta}, \mathbf{\beta})$. 
The optimization goal is to match the estimated joint position $\mathbf{J}$ via the following energy function equation:

\begin{equation}
    \label{eq:IK_mse}    
    \mathbf{E} = \mathop{\arg\min}\limits_{\bf{\theta}, \bf{\beta}} \lVert \bf{\omega} \circ (J - f(\bf{\theta}, \bf{\beta})) \rVert_2.
\end{equation}

In this equation, $\mathbf{\omega}$ represents the weight of each joint. We assign a higher weight to the fingertip joints detected as touching the screen to improve touch point accuracy. In our implementation, $\omega=2$ for touched fingertip joints and $\omega=1$ for other joints. To ensure the hand pose is plausible, the following constraints must be met:

\begin{equation}
    \label{eq:theta_range}
    \bf{\theta_{min}} \leq \bf{\theta} \leq \bf{\theta_{max}},
\end{equation}

\begin{equation}
    \label{eq:finger_height}
    \bf{f_z}(\bf{\theta}, \bf{\beta}) \geq z_0.
\end{equation}

\begin{figure*}
    \centering
    \includegraphics[width=\textwidth]{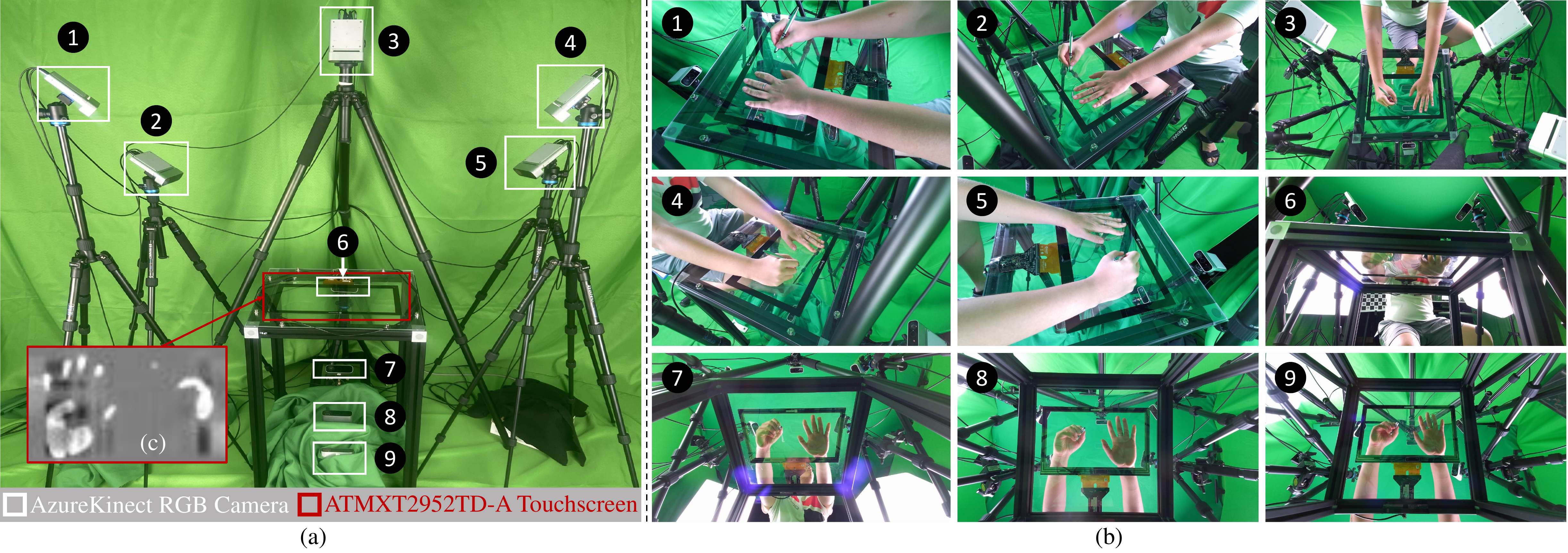}
    \caption{ The device setup of our system for capturing ground-truth hand poses and corresponding capacitive images. 
    (a) Our system consists of one capacitive touchscreen and nine RGB cameras, which are synchronized to capture RGB images of hands from different views along with the associated capacitive images. 
    (b) The nine RGB images of two hands on the touchscreen. 
    (c) The capacitive images of the two hands, captured by the touchscreen at the same time instance of the RGB images.
    }
    \label{fig:apparatus}
\end{figure*}

Here, Eq.~\ref{eq:theta_range} constrains the angle of each joint to the available motion range of the human hand, and Eq.~\ref{eq:finger_height} enforces that the fingers do not penetrate the touchscreen. $\mathbf{z_0}$ represents the joint height when touching the screen.

We employ sequential quadratic programming (SQP) to solve the constrained nonlinear optimization problem. The initial value is carried from the last frame. To maintain a consistent hand shape, we update $\beta$ for the first 20 frames when all five fingers are detected to have touched the screen. After that, we fix the hand shape by fixing $\beta$ and perform hand tracking by updating $\theta$ only.

In our implementation, we set $\mathbf{z}_0 = 5mm$ in Eq.~\ref{eq:finger_height}, allowing the generated hand mesh to penetrate the screen slightly. This design choice is based on the fact that in reality, skin deforms when fingers touch the screen. We simulate the effect of skin being flattened by the screen through post-processing, projecting the penetrating mesh vertices onto the screen. This simple method effectively mimics the skin flattening effect, providing a stronger sense of realism, which has been confirmed in lightboard applications (Sec. ~\ref{Sec:lightboard}).

\section{Dataset Construction} \label{Sec:Data Capture}

A dataset that consists of diverse and accurate hand poses and corresponding capacitive images is critical for developing and evaluating learning-based pose-tracking methods. Prior research ~\cite{DBLP:conf/uist/AhujaS021} leveraged LeapMotion for capturing 3D hand skeletons. However, the accuracy of the resulting hand gestures and finger positions is inadequate~\cite{zhang2023remotetouch}. The capacitance value of the touch screen exhibits a complex relationship with hand contact and is influenced by external factors such as environmental conditions and palm humidity, making data synthesis challenging. To tackle these challenges, we designed a data acquisition system employing multi-view stereo techniques, allowing us to obtain high-precision 3D hand pose data that is accurately aligned with the touchscreen. Based on this device setup, we have captured a dataset comprising 13.4 hours of varied hand-screen interaction clips, featuring both single and dual-hand engagements, recorded from 16 participants.

\subsection{Apparatus}

Our data capture system is shown in Fig.~\ref{fig:apparatus} (a). The touchscreen is positioned on a table with a transparent surface. To capture the hand from various directions, we employ nine cameras in our setup. We use the AzureKinect RGB camera for its high image quality and the ability to synchronize multiple cameras through cables. All sensors are connected to a single PC via USB.

The cameras are arranged to face the touchscreen from different directions, with five looking downward and four looking upward. The upward-looking cameras under the touchscreen can directly track fingers that may be occluded by the hand during gestures such as writing. The intrinsic and distortion coefficients can be accessed directly from the camera using the sensor SDK. The extrinsic is calibrated using a chessboard and the method proposed in ~\cite{zhang2000flexible}. We then capture an image of the touchscreen from each camera, manually label the four corners of the touchscreen in each image, and calculate their 3D coordinates using triangulation. This process ensures the entire system is calibrated.

To compute the accurate 3D hand pose, all nine cameras recorded images in 1080p ($1920\times1080$) resolution at 30FPS. Since the capacitive frames are captured at 15FPS, we align the RGB frames to the capacitive frames by linearly interpolating the two RGB frames that are closest to each capacitive frame according to their timestamps.

\begin{figure*}
    \centering
    \includegraphics[width=\textwidth]{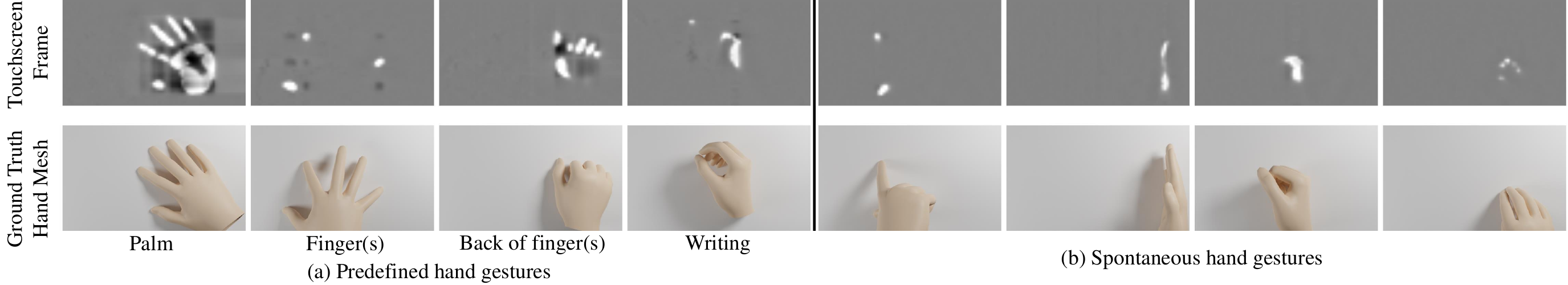}
    \caption{Depiction of the hand gestures in our dataset. During predefined gestures, participants perform hand movements within each category as shown in (a). For free hand movement, participants interchange freely among predefined categories and perform spontaneous hand gestures as shown in (b). }
    \label{fig:dataset_example}
\end{figure*}

\subsection{MVS Hand Pose Estimation}

We perform hand pose estimation from the captured multiview images using the method proposed in ~\cite{9010013}. First, we detect 2D hand joints in each frame with the hand tracking method in Mediapipe ~\cite{zhang2020mediapipe}, which can track both hands using the 21 joints model with sub-pixel accuracy. Note that although Mediapipe can predict 3D hand poses, the depth of each joint estimated from a single view is not accurate and thus is ignored in our implementation.

However, detection errors may occur, such as misclassifying left and right hands, misdetecting hands, or having significant joint estimation errors when fingers are occluded in a particular view. For robust 3D hand tracking, we need to filter these failure cases. To this end, we check the correctness of hand $h_i$ detected in one camera $i$ by checking the alignment of each joint of hand $h_j$ in another camera $j$. We average the distance between rays of the same joint from camera $i$ and $j$ by,

\begin{equation}
    \label{eq:ray_distance}
    D(h_i, h_j) = \frac{1}{21} \sum_{k=1}^{21} \left|
    (\bf{O}_i - \bf{O}_j) \cdot 
    \frac{\bf{r}_i^k \times \bf{r}_j^k}
    {\left| \bf{r}_i^k \times \bf{r}_j^k \right|} 
    \right|,
\end{equation}

Where $O_i$ and $O_j$ are optical centers of camera $i$ and $j$, $\mathbf{r}_i^k$ and $\mathbf{r}_j^k$ are rays of joint $k$ in camera $i$ and $j$. In our implementation, if $D(h_i, h_j)$ is smaller than a threshold ($1cm$ in our implementation), then the hands $h_i$ and $h_j$ are considered to be consistent in 3D space. We filter all hands that have less than 3 consistent hands from all other cameras.

Finally, for all remaining hands with high confidence, we compute their joints in 3D space via triangulation. For each joint, its 3D coordinate $\mathbf{\hat{P}}$ is calculated by minimizing the distance to rays $\mathbf{r}_i$ corresponding to the joint point in each camera $i$, as given by:

\begin{equation}
    \label{eq:ray_intersection_problem}
    \mathbf{\hat{P}} = \mathop{\arg\min}\limits_{\mathbf{P}}
    \sum_i
    \big\lVert (\mathbf{P} - \mathbf{O}_i) \times \frac{\mathbf{r}_i}{\left| \mathbf{r}_i \right|} \big\rVert _2
    ,
\end{equation}

where $\mathbf{O}_i$ represents the optical center of camera $i$. The ray $\mathbf{r}_i$ is calculated by unprojecting the detected 2D joint point on image $i$ using the intrinsic parameters and distortion coefficients of camera $i$.

\subsection{Data Acquisition and Processing}

\subsubsection{Hand Gestures}

Our dataset encompasses a set of capacitive video clips with corresponding 3D hand poses and movements that are frequently used in touch-screen-based user interactions. As shown in Fig. ~\ref{fig:dataset_example}(a), we collect four categories of pre-defined hand gestures with different hand contacts with the screen: the palm, the finger(s), the back of the fingers while in a bent position, and holding a stylus for writing. For each gesture type, the participants were guided to naturally hold the hand gesture and move the hand to traverse the entire screen at a regular pace, occasionally extending partially beyond the touch screen area. For finger gestures, the participants pressed one or more fingers on the screen and moved their hands, as well as switched their fingers on the screen in a manner similar to piano playing. For the writing gesture, the participants followed their conventional writing postures to write or draw across the entire screen. To enhance gesture diversity and ensure a smooth transition between different actions, participants were also asked to perform free hand movements, including transitioning between predefined gestures, as well as spontaneous actions, as shown in Fig.\ref{fig:dataset_example} (b). While the four categories of pre-defined hand gestures ensure the coverage of most common hand gestures, free hand movement furnishes more adaptable hand movement data and enhances the variety of hand gestures.

During the data acquisition phase, each participant was asked to perform each type of gesture with the right hand, left hand, and two hands. For the two-hand data capturing, we instructed the participants to position their hands on the screen as naturally as possible. The two hands perform the same type of gesture for pre-defined gesture capturing and both move freely in any combination for free gesture capturing. For writing with two hands, the participants performed the writing gesture with their dominant hand while placing the other hand naturally on the screen. Furthermore, we advised the participants to minimize substantial overlaps of the hands during the capturing, without deliberately avoiding minor overlaps.

\subsubsection{Participants}

Our dataset was contributed to by sixteen participants (9 males and 7 females, aged 19 to 45, mean = 28), covering a broad spectrum of hand sizes (length: 156mm to 195mm, mean = 177mm, width: 70mm to 92mm, mean = 83mm). This diversity ensures our findings are applicable to a wide user base.  

\subsubsection{Data Acquisition Strategy}

We capture hand movement data in sessions. In each session, we start with twelve clips of right-hand gestures, followed by eight clips of the left hand and eight clips of both hands, in a predetermined sequence. For the right hand, we capture 2 clips of each predefined hand gesture and 4 clips of spontaneous hand gestures. For the left and both hands, we capture 1 clip of each predefined hand gesture and the rest are spontaneous hand gestures. Given that all participants were right-handed, their right-hand gestures were more natural and flexible, we captured more right-hand data. Due to the symmetrical nature of hands, left-hand data could be effectively augmented through mirroring right-hand data in post-processing. 

For each clip, the capture takes 30 seconds of hand movement data, which contains 900 synchronized video frames from the camera array and approximately 450 frames of touchscreen data. After capturing, the system takes about 30 seconds to dump the clip data to a hard drive and participants can take a rest and get ready for the next hand gesture. For each participant, we capture 4 sessions, which take around 2 hours. 

\subsubsection{Data Processing and Filtering}

\begin{table}[ht]
    \centering
    \small
    \setlength{\tabcolsep}{2pt}  
    \begin{tabular}{c|ccccc|ccccc|ccccc}
    \toprule
    Hand & \multicolumn{5}{c|}{Right} & \multicolumn{5}{c|}{Left} & \multicolumn{5}{c}{Both} \\ \hline
    \multirow{2}{*}{Gesture} & P & F & B & W & Free & P & F & B & W & Free & P & F & B & W & Free \\
    & \multicolumn{1}{c}{117}& \multicolumn{1}{c}{118}& \multicolumn{1}{c}{117}& \multicolumn{1}{c}{115}& \multicolumn{1}{c|}{235}& \multicolumn{1}{c}{59}& \multicolumn{1}{c}{59}& \multicolumn{1}{c}{58}& \multicolumn{1}{c}{59}& \multicolumn{1}{c|}{231}& \multicolumn{1}{c}{57}& \multicolumn{1}{c}{57}& \multicolumn{1}{c}{53}& \multicolumn{1}{c}{47}& \multicolumn{1}{c}{226} \\ \hline    
    Sum     & \multicolumn{5}{c|}{702} & \multicolumn{5}{c|}{466} & \multicolumn{5}{c}{440}
    \\ \bottomrule
    \end{tabular}
    \caption{The distribution of our dataset in terms of the number of clips for each gesture, represented by Palm (P), Finger(s) (F), Back of finger(s) (B), Writing (W), and free gestures, for the right, left, and both hands.}
    \label{tab:data_statistics}
\end{table}

After data acquisition, we estimate the ground truth 3D hand poses with an offline data processing framework, which takes around 9 minutes for each clip. Quality assurance measures were stringent, with any data compromised by calibration errors, detection inaccuracies, or file corruption promptly excluded, ensuring the dataset’s integrity. 

Our data acquisition and processing pipeline culminated in a robust dataset of 1608 clips, which is approximately 13.4 hours of a variety of hand movements. The gesture distribution within our dataset is listed in Table.~\ref{tab:data_statistics}. Free hand movement occupies $43\%$ of the whole dataset, ensuring rich behavior of hand motion.

\section{Method Evaluation} \label{Sec:Evaluation}

To evaluate the effectiveness of our hand-tracking scheme, we compared the accuracy of our method with other state-of-the-art approaches, evaluated system performance with both single-hand and two-hand interaction sequences, and conducted an ablation study to verify the contribution of each system component.

\subsection{Device and System Performance} \label{sec:performance}

We utilize the Microchip ATMXT2952TD-A Evaluation Kit to acquire raw touch images. This kit features a 15.6-inch transparent touchscreen assembly, capable of capturing 8-bit $71\times41$ raw capacitive touch images at a rate of approximately 15fps and simultaneously detecting up to 16 touch points. The touchscreen is employed in both the data capture system and the whiteboard interaction prototype system.

Our hand-tracking method achieves a performance of 32fps on a PC equipped with an Intel Core i9-10980XE CPU, 64GB memory, and a Nvidia GeForce RTX 4090 GPU. For each input frame, the joint estimation network takes about 13ms to derive the joint position and the inverse kinematics takes about 18ms to infer the 3D hand pose. In our current implementation, the frame rate of the whole system is 15fps and limited by the frame rate of the capacitive imaging.

\subsection{Method Validation and Comparison}
We evaluate the accuracy and robustness of our method for various hand gestures and participants and compare our method with TouchPose ~\cite{DBLP:conf/uist/AhujaS021}, which is the state-of-the-art method for estimating hand poses from capacitive images. 

Because TouchPose's dataset in ~\cite{DBLP:conf/uist/AhujaS021} only comprises static images instead of motion sequences, it cannot be directly used for training and testing our method. Additionally, the TouchPose is designed for estimating single-hand gestures only. To facilitate a fair comparison, we construct a right-hand dataset that contains all 576 right-hand sequences in our dataset, and train and test the TouchPose and our model on this right-hand dataset. In particular, the right-hand dataset includes sequences captured from 12 participants \footnote{We exclude the data of 4 participants that consists of fewer right-hand sequences than other participants for conducting the three cross-validation evaluations.}. We subdivide the sequences of each participant into 4 sessions, each of which consists of 8 sequences of 4 predefined gesture types (two for each predefined gesture type) and 4 free movement sequences.

\paragraph{Testing Setup.}
We follow ~\cite{DBLP:conf/uist/AhujaS021} to evaluate the accuracy and generalizability of our method and compare its performance with TouchPose with the following three setups:

\begin{itemize}
    \item \textbf {P1} is a 4-fold cross-validation for evaluating the model's performance, in which three sessions of all the participants are used for model training and the remaining one session is used for testing.   
    
    \item \textbf {P2} is a 3-fold cross-validation for evaluating the model's robustness to hand gestures of new users, in which we category 12 participants into three groups and use all sequences of the two participant groups for model training and the data from the remaining group for testing. 
    
    \item \textbf {P3} is a 5-fold cross-validation for evaluating the model's robustness to new gestures, in which we train the model with the data of four out of five hand gesture types and test the model with the data of the remaining one gesture type. The free movements are counted as one gesture type.  
\end{itemize}

\noindent\paragraph{Error Metrics.}
We employed the following error metrics to measure the accuracy of the resulting 3D hand poses inferred by our method and TouchPose:

\begin{itemize}
    \item \textbf {End-point-error ($\bf{EPE}$)} computes the average Euclidean distance between the reconstructed hand joint positions and the ground truth ones, measuring the alignment accuracy of the entire hand. We further assess the error parallel to the screen surface ($\bf{EPE_{xy}}$) and the error in distance to the screen ($\bf{EPE_z}$).

    \item \textbf {End-point-error of visible finger ($\bf{EPE^v}$)} measures the end-point-error for joints of fingers that at least one joint is visible on the touchscreen. This metric is also broken down into errors in two directions, denoted by $\bf{EPE^v_{xy}}$ and $\bf{EPE^v_z}$. In remote interaction applications, $\bf{EPE^v_{xy}}$ can directly reflect the degree of alignment between the hand and the interactive content displayed on the screen.
\end{itemize}

\paragraph{Results and Analysis}
We show the results of our model and retrained TouchPose in Table \ref{tab:compare_touchpose_2}. Compared with the metric values reported in the original TouchPose paper ~\cite{DBLP:conf/uist/AhujaS021}, the hand tracking accuracy has been significantly improved by retraining using our larger and more precise dataset of hand movements. This underscores the value of our dataset in improving hand tracking accuracy.  

\begin{table}[ht]
    \centering
    \small
   \resizebox{\linewidth}{!}{
    \begin{tabular}{c|c|ccccccc} 
    \toprule
        &
        Method & 
        $\bf{EPE}$ $\downarrow$ & 
        $\bf{EPE_{xy}}$ $\downarrow$ & 
        $\bf{EPE_{z}}$ $\downarrow$ & 
        $\bf{EPE^v}$ $\downarrow$ & 
        $\bf{EPE_{xy}^v}$ $\downarrow$ & 
        $\bf{EPE_{z}^v}$ $\downarrow$ \\ 
    \midrule        
    \multirow{2}{*}{P1} & Ours & \pmb{8.85 (3.10)}  & \pmb{6.39 (2.69)} & \pmb{4.98 (1.36)} & \pmb{7.90 (2.53)} & \pmb{5.58 (2.22)} & \pmb{4.59 (1.17)}  \\
    & TouchPose & 11.3 (5.82) & 9.19 (4.75) & 5.16 (3.00) & 10.5 (5.35) & 8.56 (4.22) & 4.70 (2.99) \\
    
    \midrule 
    \multirow{2}{*}{P2} & Ours & \pmb{10.4 (3.47)} & \pmb{8.03 (2.96)} & 5.21 \pmb{(1.64)} & \pmb{8.84 (2.82)} & \pmb{6.70 (2.54)} & 4.63 \pmb{(1.25)} \\
    & TouchPose & 12.9 (8.01) & 11.0 (6.78) & \pmb{5.17} (3.62) & 11.5 (7.39) &9.90 (6.11) & \pmb{4.51} (3.61)  \\
    
    \midrule        
    \multirow{2}{*}{P3} & Ours & \pmb{11.8 (5.56)} & \pmb{9.03 (4.68)} & \pmb{5.99 (2.52)} & \pmb{9.67 (3.95)} & \pmb{7.46 (3.45)} & \pmb{4.84 (1.65)}  \\
    & TouchPose & 15.0 (8.38) & 12.5 (7.19) & 6.57 (3.67) & 13.1 (7.19) & 10.9 (6.09) & 5.59 (3.35) \\
    \bottomrule        
    \end{tabular}
    }
    \caption{Comparative analysis of hand-tracking accuracy and standard deviation (in brackets)} between our method and retrained TouchPose across different evaluation protocols (P1, P2, P3). The unit of all values in the table are millimeters (mm).
    \label{tab:compare_touchpose_2}
\end{table}

\begin{figure}
    \centering
    \includegraphics[width=\columnwidth]{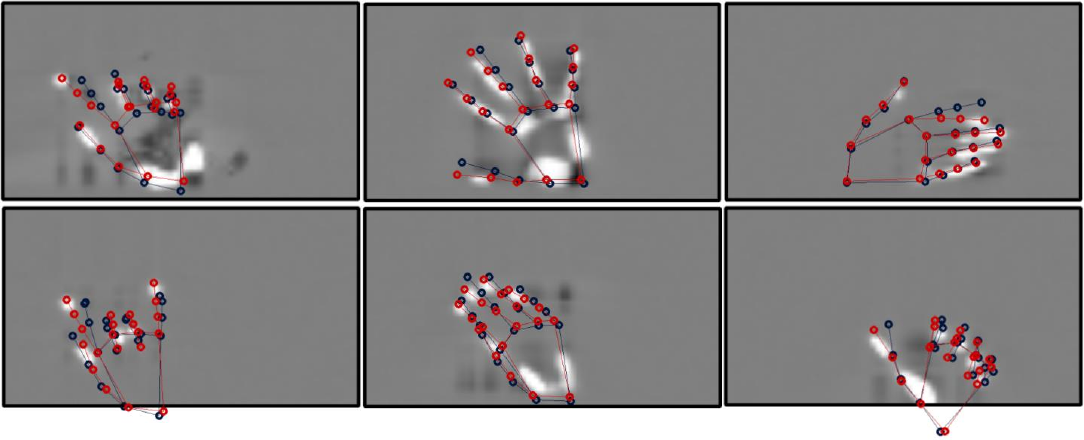}
    \caption{Visual comparison of the projected joint error $\bf{EPE_{xy}}$ between TouchPose and our method across various hand poses (best viewed on screen). By projecting the joints predicted by our method (red) onto the touch image and comparing them to TouchPose (blue), it is evident that the hand joints predicted by our method exhibit superior alignment with the touch image. }
    \label{fig:comparison_touchpose}
\end{figure}

By incorporating temporal information and employing heatmaps for joint position prediction, our method further improved hand-tracking accuracy and reduced deviation, particularly in the xy direction, which is parallel to the screen. This indicates the efficacy of precise joint prediction using heatmaps. A visual comparison of $\bf{EPE_{xy}}$ between our method and TouchPose for various gestures can be found in Fig. \ref{fig:comparison_touchpose}. For the z-direction error, our method's performance is on par with (or slightly better than) TouchPose, highlighting the inherent challenge in predicting the accurate pose of non-visible hand parts, especially the invisible fingers.  

Temporal information not only facilitates hand tracking but also aids in resolving ambiguities in hand gesture tracking. For example, when a single fingertip contacts the screen, utilizing temporal data helps accurately determine the finger's identity by tracking the alternation of different fingertips. We show this tracking improvement in the supplemental video (2:40-3:05). This kind of hand gesture ambiguity cannot be resolved by the conventional temporal filter. Our evaluation across the three protocols and comparison with TouchPose demonstrates the accuracy and generalizability of our method in hand gesture recognition.  

\subsection{More Evaluations} \label{sec:evaluation}

Our system supports interaction with the touchscreen using either one or both hands. We evaluated our full model's performance for inferring the gestures of the right, left, and both hands. An additional metric, Hand-existence-accuracy (HEA), the correct rate of hand existence, was introduced to measure the accuracy with which our method can identify the presence of either hand.  

To simulate the performance of the method in real scenarios where the user is always not the one in the training dataset, we use the data of randomly-selected 12 participants for training and the sequences of the remaining 4 participants for testing. This is similar to the P2 testing setup described above without n-fold cross-validation. The accuracy of our model for each gesture type and hand-type (left-hand, right-hand, and two-hand) are listed in Table ~\ref{tab:comprehensive_evaluation}.  

\begin{table}[t]
    \centering
    \small
    \setlength{\tabcolsep}{3pt}
    
    \begin{tabular}{c|ccccccccc} 
    \toprule
    \textbf{Hand} & \textbf{Gesture} & 
        $\bf{EPE}$  & 
        $\bf{EPE_{xy}}$  & 
        $\bf{EPE_{z}}$  & 
        $\bf{EPE^v}$  & 
        $\bf{EPE_{xy}^v}$  & 
        $\bf{EPE_{z}^v}$  & 
        $\bf{HEA}$(\%) \\    
    \midrule
    
    \multirow{5}{*}{Right} & P & 9.22 & 7.72 & 3.85 & 8.35 & 7.03 & 3.47 &  99.9 \\
    & F & 12.2 & 9.69 & 5.99 & 10.6 & 8.56 & 5.08 & 99.9 \\
    & B & 12.4 & 10.1 & 5.61 & 11.3 & 9.28 & 5.09 & 99.8 \\
    & W & 16.9 & 14.2 & 7.03 & 14.8 & 12.3 & 6.43 & 99.9 \\
    & Free & 11.7 & 9.42 & 5.47 & 10.5 & 8.58 & 4.87 & 99.8 \\
    \midrule
    \multirow{5}{*}{Left} & P & 8.52 & 6.96 & 3.82 & 7.34 & 5.98 & 3.34 & 97.9 \\
    & F & 10.6 & 8.35 & 5.32 & 9.29 & 7.21 & 4.77 & 99.8 \\
    & B & 12.4 & 9.88 & 5.97 & 11.3 & 9.00 & 5.45 & 99.1 \\
    & W & 14.1 & 10.8 & 7.49 & 9.77 & 7.59 & 5.01 & 99.4 \\
    & Free & 11.4 & 9.03 & 5.60 & 10.3 & 8.19 & 5.08 & 99.3 \\
    \midrule
    \multirow{5}{*}{Both} & P & 10.8 & 8.99 & 4.57 & 9.59 & 7.99 & 4.08 & 100\\
    & F & 12.1 & 9.52 & 6.02 & 10.1 & 8.17 & 4.91 & 100 \\
    & B & 12.5 & 10.1 & 5.66 & 11.9 & 9.76 & 5.31 & 100 \\
    & W & 14.0 & 11.4 & 6.27 & 11.7 & 9.58 & 5.16 & 100 \\
    & Free & 11.5 & 9.33 & 5.40 & 10.6 & 8.65 & 4.85 & 100 \\
    \midrule
    Avg &  & 11.8 & 9.46 & 5.58 & 10.4 & 8.35 & 4.86 & 99.7 \\ 
    
    \bottomrule
    \end{tabular}
    \caption{Evaluation of hand-tracking performance across different hands and gestures (P: Palm, F: Finger(s), B: Back of finger(s), W: Writing). The unit of all values for $\bf{EPE}$ are millimeters (mm).}
    \label{tab:comprehensive_evaluation}
\end{table}

\begin{figure*}
    \centering
    \includegraphics[width=\textwidth]{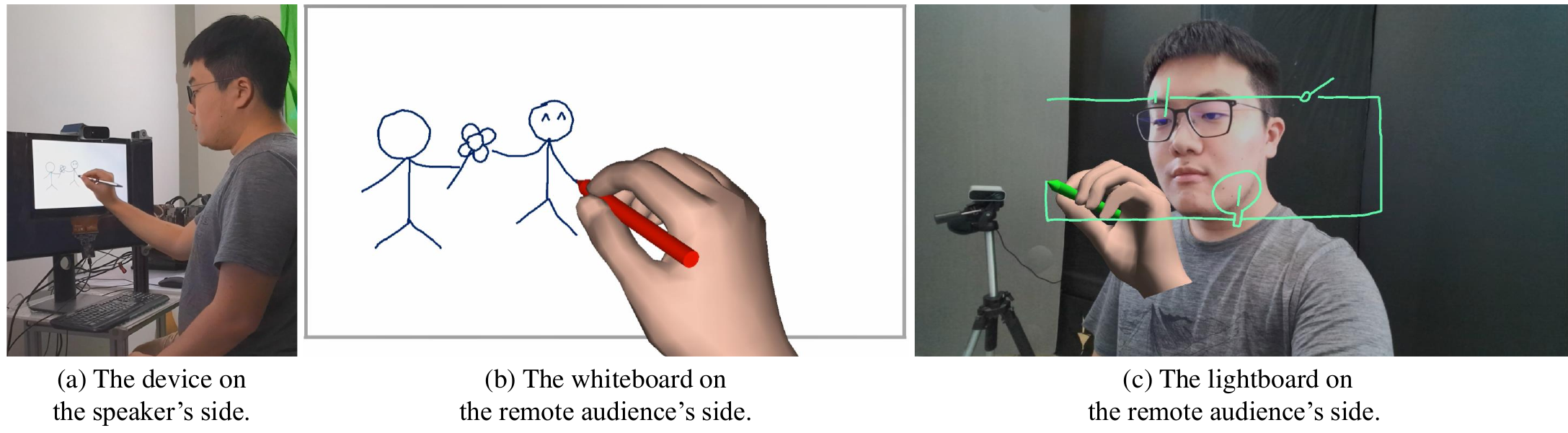}
    \caption{The two remote communication applications for method evaluation.  (a) The device setup on the speaker's side. (b) A screenshot of the whiteboard application, where both remote audiences and the speaker are positioned on the same side of a virtual whiteboard. (c) A screenshot of the lightboard application where the speaker and remote audiences are on opposite sides of a semi-transparent virtual whiteboard, enabling audiences to observe the speaker's expressions and interactions with the whiteboard. }
    \label{fig:application}
\end{figure*}

The system achieves an overall accuracy of $11.8mm$. For fingers in contact with the touchscreen surface, the accuracy reached $10.4mm$, with $\bf{EPE^v_{xy}}$=$8.35mm$ (error of contacted fingers in the direction parallel to the screen surface). As a result, the shifting of the rendered fingers is hardly noticeable in our remote whiteboard applications.

The accuracy of hand tracking varies with different gestures and the contact area with the screen. Generally, the larger the contact area that can display the overall posture of the hand, the smaller the error. For example, when the palm touches the screen (i.e. the gesture P), the fingers are spread out on the screen, and the capacitive image can reflect the overall shape of the palm, resulting in small errors in both horizontal and vertical directions. When only part of the fingertips or the back of the fingers contacts the screen (i.e. gesture F and B), the posture of the palm and the lift up fingers need to be estimated, leading to a decrease in hand tracking accuracy. For writing, gestures among individuals exhibit variability, while the capacitive image only captures the palm's edge. Consequently, our method is difficult to accurately reconstruct the specific gesture of holding a pen based solely on the capacitive image. 

In terms of end-point-error metric, a consistent trend for accuracy across various hand gestures is observed for both single-hand and dual-hand scenarios, with the left hand's accuracy marginally exceeding that of the right hand. This discrepancy can be attributed to the fact that all participants in our dataset are right-handed, which results in a reduced range of motion for the left hand due to its lesser flexibility.

In terms of the accuracy of determining hand presence, the judgment for both hands is very stable when both are in contact with the screen. However, when only one hand touches the screen, it becomes necessary to identify which hand it is, and the success rate of recognizing the right hand is slightly higher than that of the left hand. This may be because the dataset contains more data for the right hand than for the left hand, thus allowing for more accurate recognition of right-hand movements. These results also indicate that in future work, our dataset should include data from participants who are left-handed, to achieve a balance in the quantity of data and the range of motion for gestures between the left and right hands.

\subsection{Ablation Study}

\begin{table}[t]
    \centering
    \small

    \begin{tabular}{l|ccc} 
    \toprule
    Method & $\bf{EPE}$(mm) $\downarrow$ & $\bf{HEA}$(\%) $\uparrow$ \\ 
    \midrule
    W/o Temporal info (GRU) & 13.9 & 96.3 \\
    W/o SE block & 12.9 & 99.6 \\
    W/o test-time augmentation & 13.0 & 99.5 \\
    Our full model  & \textbf{11.8} & \textbf{99.7} \\

    \bottomrule
    \end{tabular}
    \caption{The results of the ablation study. We evaluate the contribution of each module to the accuracy of the estimated hand pose.}
    \label{tab:Ablation}
\end{table}

We conduct an ablation study by removing each component in our method and retrain the model with the setup described in Section \ref{sec:evaluation}. We then follow the scheme in Section \ref{sec:evaluation} to evaluate each retrained model on testing sequences of 4 participants. As shown in Table ~\ref{tab:Ablation}, each component makes an important contribution to the accuracy of the estimated hand poses. Combining all components, our full model achieves the highest overall performance, highlighting the synergistic effect of the integrated system design.

\section{Applications} \label{Sec:Real-Time Implementation}

In this section, we show potential applications of our method in remote whiteboard interactions and the user experience of our system. Our prototype whiteboard system setup, as illustrated in Fig.~\ref{fig:application} (a), follows a typical tablet configuration, complete with a touchscreen and a frontal camera. We attach the touchscreen to the monitor screen and adjust the display viewport to align with the touchscreen boundaries. An RGB camera is positioned at the border of the touchscreen and serves as the frontal camera.

\subsection{Whiteboard}

We developed a whiteboard prototype system on the hardware device depicted in Fig.\ref{fig:application} (a). On the speaker's side, the experience is similar to that of a traditional whiteboard system, allowing them to write directly on the screen using a stylus. On the audience's side, not only is the whiteboard content displayed, but also the speaker's hand and pen are visible, as illustrated in Fig.\ref{fig:application} (b).

Utilizing the touchscreen's ability to simultaneously capture touchpoint coordinates and the capacitive touch image, we calculate the stylus tip's position based on touchpoint coordinates and estimate the hand posture holding the pen using the contact image. Subsequently, we fit a 3D pen model to the hand posture, aligning the pen tip with the touchpoint. This method enables the audience to observe the speaker's writing process on the whiteboard, achieving a visual presenter-like effect.

Our system can estimate the gestures of both hands simultaneously, allowing the speaker to perform operations directly through gestures, such as erasing using a fist. For additional whiteboard interaction examples, please refer to our accompanying video. Our whiteboard system not only provides convenience for speakers but also enhances the visual experience for the audience, making it a valuable asset for presentations and discussions.

\subsection{Lightboard} \label{Sec:lightboard}

The lightboard application emulates the experience of a speaker and remote audiences situated on opposite sides of a semi-transparent whiteboard, enabling audiences to view the speaker's expressions and interactions with the written content. Conventional lightboards require specialized equipment, but our hand-tracking method on touchscreens enables us to easily create a lightboard experience on touchscreen devices.

In our system, the lightboard employs the same device as the whiteboard mentioned earlier, but utilizes the RGB camera to capture live video of the speaker. Strokes, hands, and the pen are rendered from the backside of the screen, with the video displayed as the background. In this way, the speaker appears to be writing on a transparent glass surface, as illustrated in Fig.\ref{fig:application} (c). Notice that when the speaker draws with the stylus on the touchscreen, his eyes are also concentrated on the drawing point. The consistency of hand movements and gaze enhances the speaker's sense of realism and increases the audience's engagement.

\begin{figure}
    \centering
    \includegraphics[width=\columnwidth]{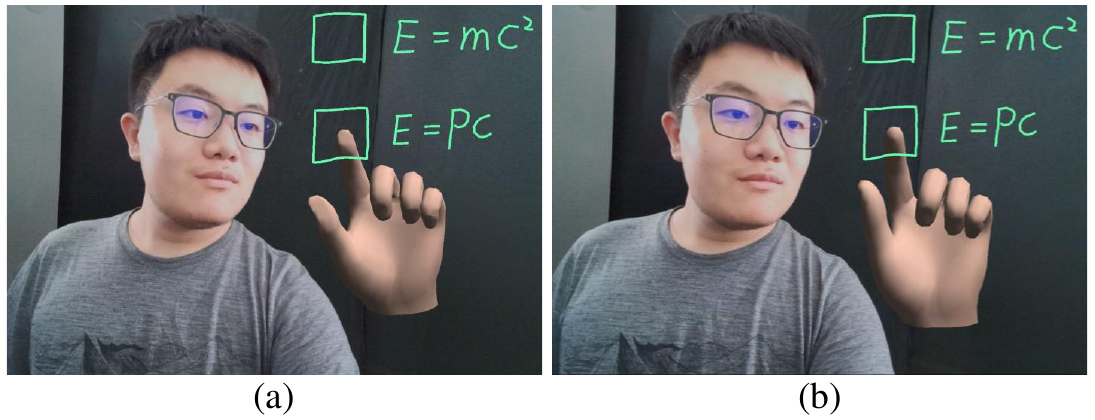}
    \caption{
    Passive hand deformation caused by touch pressure. 
    (a) The rendering of the hand featuring deformed fingertips, simulated by our constrained IK solver. 
    (b) The rendering of the hand without any fingertip deformation.
    }
    \label{fig:lightboard_cmp}
\end{figure}

To further enhance the visual experience of hand interaction on the lightboard, we employ passive deformation of the hand, enabled by the constrained IK solver. By deforming regions of the hand mesh that touch the screen, the flattened area delivers visual cues that the hand is touching the screen, such as the fingertips, as shown in Fig.\ref{fig:lightboard_cmp} (a), and the little finger, as depicted in Fig.\ref{fig:application} (c). 

Given that the content on the glass appears mirrored when observed from the back, we implement a horizontal flip on the synthesized video before displaying it to audiences. This approach is a common solution employed by lightboard devices. Consequently, the displayed content appears normal, while the speaker is flipped and appears to be writing with his left hand (assuming the speaker is right-handed). According to the user experience tests, this mirroring does not negatively impact the overall user experience, maintaining a seamless and professional appearance.

\subsection{Other Applications} \label{sec:gesture_interaction}

\begin{figure}
    \centering
    \includegraphics[width=\columnwidth]{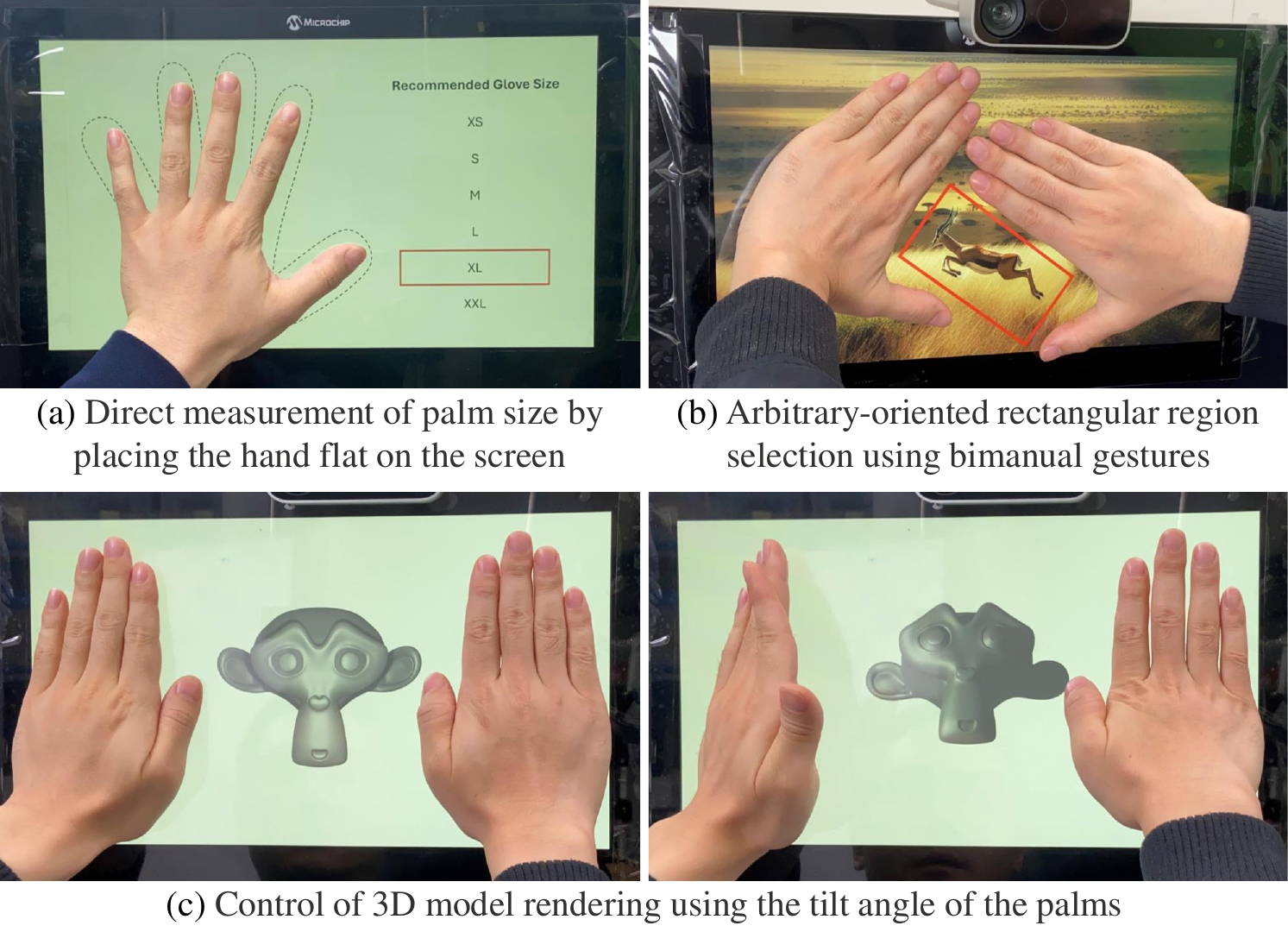}
    \caption{Illustrative examples of multimodal gestural interactions using our hand pose estimation method.}
    \label{fig:application2}
\end{figure}

Aside from whiteboard applications, our method's capability for real-time estimation of dual-hand postures on touchscreens unveils several more potential applications. In Fig. ~\ref{fig:application2} (a), we showcase a potential application in e-commerce: simply by placing the palm flat against the touchscreen, one can directly estimate the size of the palm, facilitating the selection of appropriately sized gloves. Additionally, our method enables exceptionally convenient user-computer interactions through gestures. Illustrated in Fig. ~\ref{fig:application2} (b) is the capability to directly delineate a rectangular area of any orientation on the screen using the thumb and index finger of both hands; and in Fig. ~\ref{fig:application2} (c), the capability to manipulate 3D model rendering by altering the palm-to-touchscreen contact angle. The left and right hands independently modulate the model's lighting and rotation, respectively, simplifying the process of examining 3D models for users. We are confident that touchscreens with the ability for real-time gesture estimation will unlock further possibilities in human-computer interaction.

\subsection{User Experience} \label{sec:user_experience}

To validate the effectiveness of our approach in enhancing whiteboard and lightboard applications, we conducted an informal user experience test with 16 participants. Before the test, we prepared four video clips showcasing four scenarios: whiteboard and lightboard, with and without hand display.  The participants, acting as the audience, compared the hand display modes in each scenario to determine which approach better facilitated their focus on the content written by the presenter. In the whiteboard mode, 13 participants (81\%) found that incorporating the hand display improved their attention, while in the lightboard mode, 15 participants (94\%) agreed.

In the lightboard mode, we asked participants if they noticed that the displayed video was mirrored, with the speaker appearing to write with their left hand. Only two participants (12\%) observed this independently, and they all agreed that the mirroring had no impact on their experience. Comparing the conditions shown with and without passive hand deformation, all participants agreed that passive hand deformation led them to believe that the hand had touched the screen.

For our current hand display implementation, we utilized a non-photorealistic rendering technique. We also presented an alternative approach where hand images were captured directly through a camera and transmitted in real time to the interactive whiteboard. Among participants, eight (50\%) preferred the non-photorealistic rendering for reasons such as privacy protection, consistency with the whiteboard drawing style, and simplicity of texture, which helped them concentrate on the content on the whiteboard. All participants agreed that if a photorealistic hand drawing was not available, our current display method serves as a satisfactory alternative for fulfilling interactive requirements.

In conclusion, the user experience test results demonstrate that our hand-tracking technique significantly enhances participants' attention and engagement in both whiteboard and lightboard modes.

\section{Limitations and Future Work} \label{Sec:Conclusion}

Our designed gestures cover common gestures associated with touchscreen interactions and allow participants to move freely as a supplement, yet it still cannot encompass all possible gestures. Therefore, it fails to accurately track extreme gestures. Currently, our data collection system requires post-processing to calculate the 3D skeleton of the hand. Our hand motion dataset, which captured 13.4 hours of hand motion data, required 32 hours for filming and 250 hours for post-processing. This process does not allow for the timely identification of issues during filming, leading to wasted data capture efforts and low data collection efficiency. 
To address this issue, we plan to use more gesture categories (similar to ~\cite{DBLP:conf/mhci/ChoiM021}) to guide users in capturing hand motion data, and use a real-time hand tracking system based on vision ~\cite{yu2022synchronized} or motion capture systems ~\cite{le2018infinitouch} for real-time quality inspection.
This would significantly improve data collection efficiency, making it possible to construct large-scale datasets that include more participants and a greater diversity of gestures.

Our current inverse kinematics solver does not adequately consider scenarios where the hand is holding an object, such as a pen. This approach, which merely positions the pen directly at the hand's location, often results in an unnatural pen-holding posture. Furthermore, we constrain the pen tip's position to the touch points on the screen, leading to noticeable jittering when the pen tip is lifted. For future enhancements, we propose integrating the pen into the kinematic model to more accurately represent gestures involving objects. This modification is expected to produce more natural hand movements, particularly when interacting with items like a pen.

From an application perspective, our current system only displays cartoon-style hands, without showing arms. Investigating the challenge of rendering realistic hand and arm displays, and achieving consistency with the user's body in a lightboard scene, presents a compelling research direction. Our existing whiteboard system operates in a broadcast mode, preventing other users from participating in the interaction. Exploring the design of touchscreen-based whiteboard and lightboard systems that support multi-user participation is a valuable endeavor. Additionally, at present, we simulate the tablet experience by affixing a touchscreen development kit to the screen. We plan to customize the device driver to enable our system to operate on actual tablets and mobile phones.

\section{Conclusion} \label{Sec:Conclusion}

We present a real-time 3D hand-tracking method for estimating hand poses from capacitive touch images. Our system utilizes a Recurrent U-Net network structure to process capacitive images and deliver temporally consistent and accurate hand poses while two hands interact with the screen. We also captured a ground-truth hand movement dataset via a multi-view acquisition system, which is used for network training and improves the pose estimation performance. Our method greatly improves the generality and accuracy of capacitive image-based pose estimation and enhances the user experience of whiteboard-based remote communication systems. We believe that our method could help facilitate more intuitive and seamless human interaction with touchscreens.

\begin{acks}
We thank Zhiqi Li for early exploration of the project, Yu Liu, Sicheng Xu, Chong Li, Guojun Chen for their help in data capturing, Yang Liu for discussion, and our users for their valuable feedback. We also thank the anonymous reviewers for their valuable suggestions. 
\end{acks}

\bibliographystyle{ACM-Reference-Format}
\bibliography{sample-base}

\appendix

\section{Appendix}

In the Appendix, we provide some details of the modules used in our network.

\begin{figure}
    \centering
    \includegraphics[width=0.8\columnwidth]{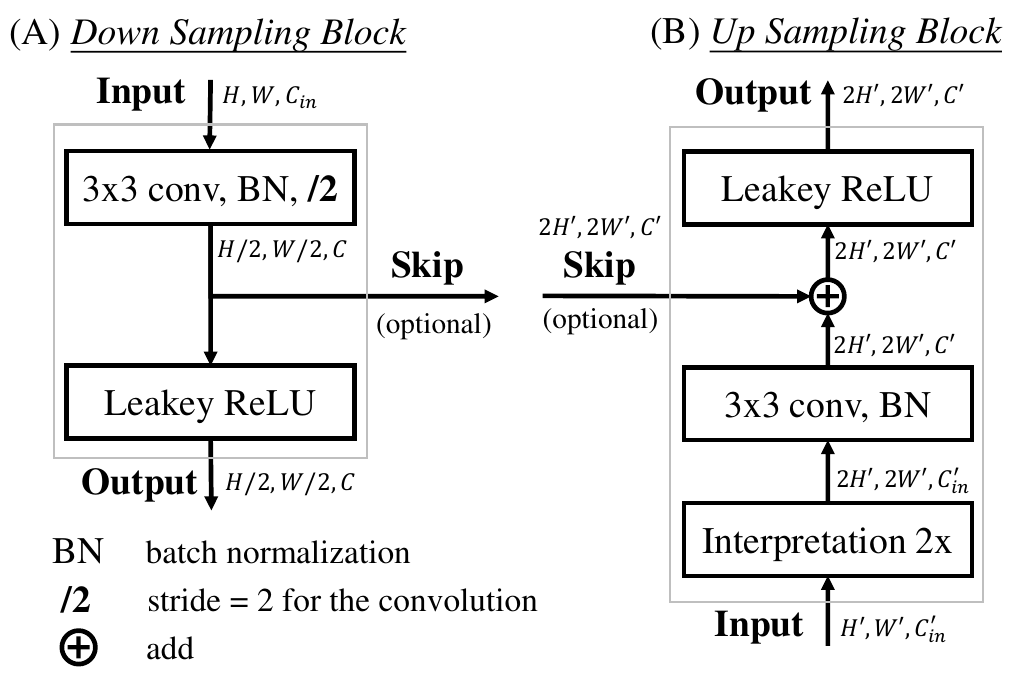}
    \caption{Our down sampling block and up sampling block.}
    \label{fig:UpDownBlocks}
\end{figure}

Fig.~\ref{fig:UpDownBlocks} illustrates the structure of the downsampling and upsampling blocks. The downsampling block commences with a 2D convolution layer featuring a kernel size of 3x3 and a stride of 2, which is then followed by batch normalization and a leaky ReLU activation. Conversely, the upsampling block performs the inverse process. Blocks at the same level are interconnected through skip connections, as depicted in Fig.~\ref{fig:UpDownBlocks}.

\begin{figure}
    \centering
    \includegraphics[width=0.4\columnwidth]{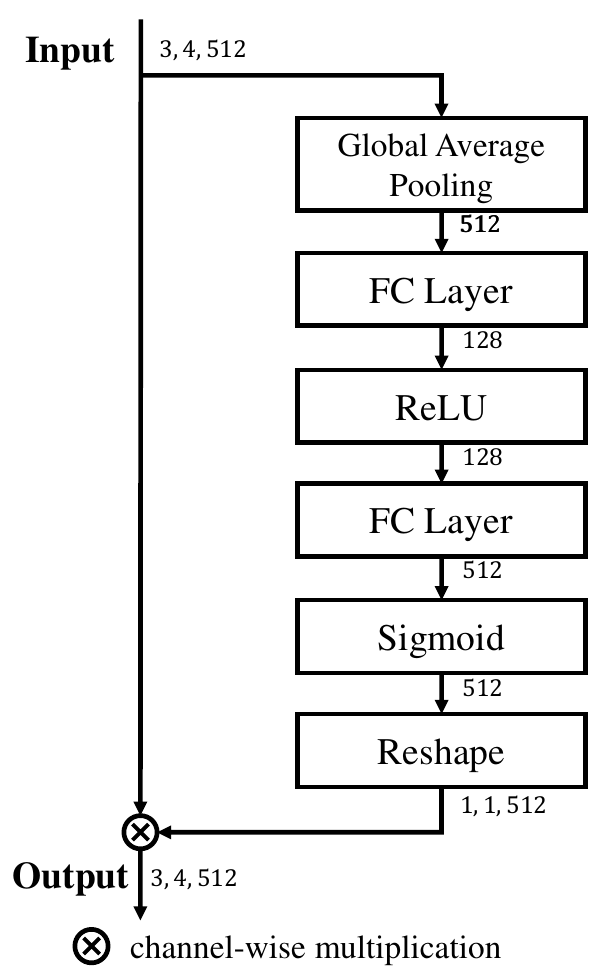}
    \caption{Illustration of the SE-block in our model.}
    \label{fig:SE-Block}
\end{figure}

Upon completing the downsampling process five times, we obtain a $4\times3\times512$ latent code that is subsequently passed through a Squeeze-and-Excitation (SE) block, as illustrated in Fig.\ref{fig:SE-Block}. Adhering to the standard implementation of the SE block as described in \cite{DBLP:conf/cvpr/HuSS18}, this method has been demonstrated to enhance performance through our ablation study.

\end{document}